\documentclass[10pt, final,narrowdisplay]{elsart}

\usepackage{graphicx}
\usepackage{anysize} 

\usepackage{amssymb}
\usepackage{amsmath}

\begin{document}

\newcommand{\HALF}{\frac{1}{2}}
\newcommand{\pd}[2]{\frac{\partial #1}{\partial #2}}
\renewcommand{\vec}[1]{\mathbf{#1}}
\begin{frontmatter}



\title{A Simple and Accurate Riemann Solver for Isothermal MHD}


\author[a,b]{A. Mignone}
\ead{mignone@ph.unito.it}
\address[a]{Dipartimento di Fisica Generale ``Amedeo Avogadro", 
         Universit\`a degli Studi di Torino, 
         via Pietro Giuria 1,
         10125 Torino, Italy}
\address[b]{INAF/Osservatorio Astronomico di Torino,
                 Strada Osservatorio 20, 
                 10025 Pino Torinese, Italy}

\begin{abstract}
 A new approximate Riemann solver for the equations of 
 magnetohydrodynamics (MHD) with an isothermal equation of state is presented. 
 The proposed method of solution draws on the recent work of 
 Miyoshi and Kusano, in the context of adiabatic MHD, where an 
 approximate solution to the Riemann problem is sought in terms of 
 an average constant velocity and total pressure across the Riemann fan.
 This allows the formation of four intermediate states enclosed by two outermost 
 fast discontinuities and separated by two rotational waves and an entropy mode.
 In the present work, a corresponding derivation for the isothermal
 MHD equations is presented. 
 It is found that the absence of the entropy mode leads to a different 
 formulation which is based on a three-state representation rather than four.
 Numerical tests in one and two dimensions demonstrates that the new solver
 is robust and comparable in accuracy to the more expensive linearized solver
 of Roe, although considerably faster.   
\end{abstract}

\end{frontmatter}

\section{Introduction}
\label{sec:intro}
%
%
%

The numerical solution of the magnetohydrodynamics (MHD) equations 
has received an increasing amount of attention over the last few decades 
motivated by the growing interests in modeling highly nonlinear flows.
Numerical algorithms based on upwind differencing techniques have 
established stable and robust frameworks for their ability to  
capture the dynamics of strong discontinuities. 
This success owes to the proper computation of numerical fluxes
at cell boundaries, based on the solution of one-dimensional 
Riemann problems between adjacent discontinuous states.
The accuracy and strength of a Riemann solver is intrinsically tied to 
the degree of approximation used in capturing the correct wave pattern
solution. 

In the case of adiabatic flows, an arbitrary initial discontinuity evolves 
into a pattern of constant states separated by seven waves, 
two pairs of magneto-sonic waves (fast and slow),
two Alfv{\'e}n discontinuities and one contact or entropy mode.
With the exception of the entropy mode, a wave separating two 
adjacent states may be either a shock or a rarefaction wave.
In the limit of efficient radiative cooling processes or 
highly conducting plasma, on the other hand, the adiabatic assumption 
becomes inadequate and the approximation of an isothermal flow is better 
suited to describe the flow.
In this case, the solution of the Riemann problem is similar 
to the adiabatic case, with the exception of the 
contact mode which is absent.

Although analytical solutions can be found with a high degree of accuracy
using exact Riemann solvers, the resulting numerical codes are
time-consuming and this has pushed researcher's attention toward other 
more efficient strategies of solution.
From this perspective, a flourishing number of approaches has been developed in 
the context of adiabatic MHD, see for example 
\cite{DW, RJ95, CG97, Balsara} and references therein.
Similar achievements have been obtained for the isothermal MHD 
equations, see \cite{Balsara, KRJH99}.

Recently, Miyoshi and Kusano (\cite{MK05}, MK henceforth) proposed a 
multi-state Harten-Lax-van Leer (HLL, \cite{HLL83}) Riemann solver for 
adiabatic MHD. The proposed ``HLLD" solver relies   
on the approximate assumption of constant velocity and total pressure over
the Riemann fan. This naturally leads to a four-state representation of
the solution where fast, rotational and entropy waves are allowed to form.
The resulting scheme was shown to be robust and accurate.
The purpose of the present work is to extend the approach developed 
by MK to the equations of isothermal magnetohydrodynamics (IMHD henceforth). 
Although adiabatic codes with specific heat ratios close to one 
may closely emulate the isothermal behavior, it is nevertheless advisable 
to develop codes specifically designed for IMHD, because of the greater ease of 
implementation and efficiency over adiabatic ones. 

The paper is structured as follows: in \S\ref{sec:eqns} the equations of
IMHD are given; in \S\ref{sec:solver} the new Riemann solver is derived. 
Numerical tests are presented in \S\ref{sec:num} and conclusions are
drawn in \S\ref{sec:conclusions}.


\section{Governing Equations}
\label{sec:eqns}
%
%
%

The ideal MHD equations describing an electrically conducting perfect
fluid are expressed by conservation of mass,
\begin{equation}\label{eq:continuity}
 \pd{\rho}{t} + \nabla\cdot(\rho\vec{v}) = 0 \,,
\end{equation}
momentum,
\begin{equation}\label{eq:momentum}
 \pd{\vec{m}}{t} + \nabla\cdot\left(\vec{m}\vec{v} - \vec{B}\vec{B} + \vec{I}p_T\right) = 0\,,
\end{equation}
and magnetic flux,
\begin{equation}\label{eq:induction}
 \pd{\vec{B}}{t} - \nabla\times\left(\vec{v}\times\vec{B}\right) = 0\,.
\end{equation}
The system of equations is complemented by the additional 
solenoidal constraint $\nabla\cdot\vec{B} = 0$ and by an equation 
of state relating pressure, internal energy and density and
Here $\rho$, $\vec{v}$ and $\vec{m}=\rho\vec{v}$ are used to denote,
respectively, density, velocity and momentum density.
The total pressure, $p_T$, includes thermal ($p$) and magnetic
($|\vec{B}|^2/2$) contributions (a factor of $\sqrt{4\pi}$ has been
reabsorbed in the definition of $\vec{B}$).
No energy equation is present since I will consider the isothermal 
limit for which one has $p = a^2\rho$, where 
$a$ is the (constant) isothermal speed of sound.
With this assumption the total pressure can be written as:
\begin{equation}
 p_T = a^2\rho + \frac{|\vec{B}|^2}{2} \,.
\end{equation}

Finite volume numerical schemes aimed to solved
(\ref{eq:continuity})--(\ref{eq:induction}) rely
on the solution of one-dimensional Riemann problems 
between discontinuous left and right states at zone edges.
Hence without loss of generality, I will focus, in what follows, 
on the one-dimensional hyperbolic conservation law
\begin{equation}\label{eq:oned}
 \frac{\partial\vec{U}}{\partial t} 
  + 
 \frac{\partial\vec{F}}{\partial x} = 0 \,,
\end{equation}
where the vector of conservative variables $\vec{U}$ and
the corresponding fluxes $\vec{F}$ are given by
\begin{equation} 
\vec{U} = \left(\begin{array}{c}
 \rho   \\ \noalign{\medskip}
 \rho u \\ \noalign{\medskip}
 \rho v \\ \noalign{\medskip}
 \rho w \\ \noalign{\medskip}
 B_y    \\ \noalign{\medskip}
 B_z    \end{array}\right)
 \,,\quad
\vec{F} = \left(\begin{array}{c}
 \rho u \\ \noalign{\medskip}
 \rho u^2 + p_T - B_x^2 \\ \noalign{\medskip}
 \rho vu  - B_xB_y   \\ \noalign{\medskip}
 \rho wu  - B_xB_z   \\ \noalign{\medskip}
 B_yu - B_xv    \\ \noalign{\medskip}
 B_zu - B_xw   \end{array}\right) \,,
\end{equation}
where $u,v$ and $w$ are the three components of 
velocity in the $x$, $y$ and $z$ directions, respectively.
The solution of Eq. (\ref{eq:oned}) may be achieved through 
the standard two-point finite difference scheme
\begin{equation}\label{eq:1st_ord}
 \vec{U}_i^{n+1} = \vec{U}_i^n - \frac{\Delta t}{\Delta x}
                   \left(\hat{\vec{F}}_{i+\HALF} - \hat{\vec{F}}_{i-\HALF}\right)\,,
\end{equation}
where $\hat{\vec{F}}_{i\pm\HALF}$ is the numerical flux function,
properly computed by solving a Riemann problem between 
$\vec{U}_i$ and $\vec{U}_{i\pm1}$.
   
As already mentioned, the solution to the Riemann problem 
in IMHD consists of a six wave pattern, with two pairs of 
magneto-sonic waves (fast and slow) and two Alfv{\'e}n (rotational) waves.
These families of waves propagate information with 
characteristic signal velocities given by the eigenvalues 
of the Jacobian $\partial\vec{F}(\vec{U})/\partial\vec{U}$:
\begin{equation}\label{eq:eigenvalues}
 \lambda_{f_\pm} = u \pm c_f \,,\quad
 \lambda_{s_\pm} = u \pm c_s \,,\quad
 \lambda_{a_\pm} = u \pm c_a \,,\quad
\end{equation}
where the fast, Alfv{\'e}n and slow velocity are defined as 
in adiabatic MHD:
\begin{equation} 
 c_f = \left[\frac{1}{2}\left(a^2 + \vec{b}^2 + 
       \sqrt{\left(a^2 + \vec{b}^2\right)^2 - 4a^2b_x^2}\right)\right]^{1/2} \,,
\end{equation}
\begin{equation}
 c_a  = |b_x|
\end{equation}
\begin{equation}
 c_s = \left[\frac{1}{2}\left(a^2 + \vec{b}^2 -
       \sqrt{\left(a^2 + \vec{b}^2\right)^2 - 4a^2b_x^2}\right)\right]^{1/2} \,,
\end{equation}
with $\vec{b} = \vec{B}/\sqrt{\rho}$.

The contact or entropy mode is absent.
Fast and slow waves can be either shocks, where flow quantities experience 
a discontinuous jump, or rarefaction waves, characterized by a smooth 
transition of the state variables.
Across the rotational waves, similarly to the adiabatic case, 
density and normal component of velocity are continuous, 
whereas tangential components are not.
In virtue of the solenoidal constraint, the normal component of the magnetic 
field is constant everywhere and should be regarded as a given parameter.

\section{The HLLD Approximate Riemann Solver}
\label{sec:solver}
%
%
%

\begin{figure}\begin{center}
 \includegraphics*[width=0.75\textwidth]{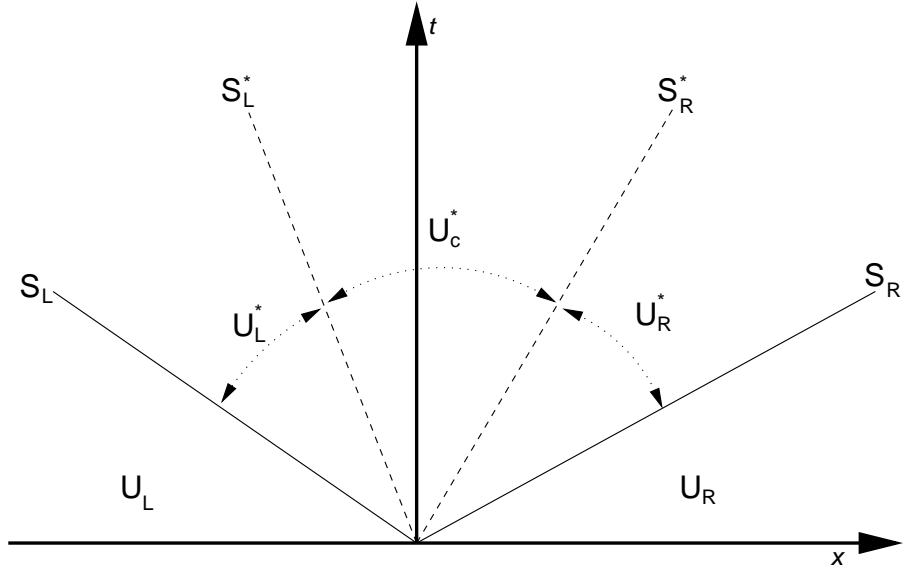}
 \caption{Schematic representation of the approximated wave structure 
          used to describe the Riemann fan. The initial states $U_L$ and $U_R$ 
          are connected to each other through a set of four waves ($S_L$, $S^*_L$
          $S^*_R$ and $S_R$) across which some or all of the fluid variables
          can be discontinuous. The four discontinuities isolate three intermediate 
          constant states $U^*_L$, $U^*_c$ and $U^*_R$ which correspond 
          to the approximate solution of the Riemann problem.}
 \label{fig:fan}
\end{center}\end{figure}

In what follows, I will assume that the solution to the isothermal 
MHD Riemann problem can be represented in terms of three constant 
states, $\vec{U}^*_L$, $\vec{U}^*_c$ and $\vec{U}^*_R$, separated by four waves: 
two outermost fast magneto-sonic disturbances with speeds
$S_L$ and $S_R$ ($S_L < S_R$) and two
innermost rotational discontinuities with velocities
$S^*_L$ and $S^*_R$, see Fig. \ref{fig:fan}.
Slow modes cannot be generated inside the solution.
Since no entropy mode is present and rotational discontinuities carry jumps in 
the tangential components of vector fields only, it follows that one can regard 
density, normal velocity and total pressure as constants over the fan.
This assumption is essentially the same one suggested by MK but it leads,
in the present context, to a representation in terms of three states rather 
than four.
Quantities ahead and behind each wave must be related
by the jump conditions
\begin{equation}\label{eq:out}
 S_\alpha\left(\vec{U}^*_\alpha - \vec{U}_\alpha\right) = 
 \vec{F}^*_\alpha - \vec{F}_\alpha \,,
\end{equation}
across the outermost fast waves, and
\begin{equation}\label{eq:in}
 S^*_\alpha\left(\vec{U}^*_c - \vec{U}^*_\alpha\right) = 
 \vec{F}^*_c - \vec{F}^*_\alpha  \,,
\end{equation}
across the innermost Alfv{\'e}n waves. Here $\alpha = L$ or $\alpha=R$ 
is used for the left or right state, respectively.
Note that $\vec{U}^*$ and $\vec{F}^*$ are both unknowns 
of the problem and $\vec{F}^*$ should not be confused with 
$\vec{F}(\vec{U}^*)$, although one always has 
$\vec{F}_\alpha = \vec{F}(\vec{U}_\alpha)$.

The solution to the problem must satisfy the consistency 
condition which, in terms of $\vec{U}^*$, can be obtained 
by direct summation of Eq. (\ref{eq:out}) and (\ref{eq:in})
across all waves:
\begin{equation}\label{eq:consistency1}
 \frac{  \left(S_L^* - S_L\right)  \vec{U}^*_L 
       + \left(S_R^* - S_L^*\right)\vec{U}^*_c
       + \left(S_R   - S_R^*\right)\vec{U}^*_R}
      {S_R - S_L} = \vec{U}^{\rm hll}  \,,
\end{equation} 
where 
\begin{equation}\label{eq:uhll}
 \vec{U}^{\rm hll} = \frac{S_R\vec{U}_R - S_L\vec{U}_L - \vec{F}_R + \vec{F}_L}
                          {S_R - S_L}  
\end{equation}
is the HLL single average state.
Likewise, after dividing Eq. (\ref{eq:out}) and (\ref{eq:in}) by 
$S_\alpha$ and $S^*_\alpha$ (respectively) and adding 
the resulting equations, the following condition for the fluxes $\vec{F}^*$ is found:
\begin{equation}\label{eq:consistency2}
 \frac{   \left(\frac{1}{S^*_L} - \frac{1}{S_L}\right)\vec{F}^*_L 
        + \left(\frac{1}{S_R^*} - \frac{1}{S_L^*}\right)\vec{F}^*_c 
        + \left(\frac{1}{S_L^*} - \frac{1}{S_R^*}\right)\vec{F}^*_R}
      {\frac{1}{S_R} - \frac{1}{S_L}} = \vec{F}^{\rm hll}  \,,
\end{equation} 
where
\begin{equation}\label{eq:fhll}
 \vec{F}^{\rm hll} = \frac{S_R\vec{F}_L - S_L\vec{F}_R + S_RS_L(\vec{U}_R - \vec{U}_L)}
                          {S_R - S_L}
\end{equation}
is the HLL single average flux. 
Equation (\ref{eq:fhll}) represents the average flux function inside
the Riemann fan and can be considered as the ``dual" relation 
of (\ref{eq:uhll}). Indeed, Eq. (\ref{eq:fhll}) could have been obtained by 
letting $\vec{U}_\alpha\to\vec{F}_\alpha$, 
$\vec{F}_\alpha\to\vec{U}_\alpha$ and $S_\alpha\to 1/S_\alpha$ 
on the right hand side of Eq. (\ref{eq:uhll}).
 
In this approximation, density, normal velocity and total pressure 
do not experience jumps across the inner waves 
and are constant throughout the Riemann fan, i.e., 
$\rho^*_L = \rho^*_c = \rho^*_R \equiv\rho^*$,
$u^*_L = u^*_c = u^*_R \equiv u^*$, 
$p_{T_L}^* = p_{T_c}^* = p_{T_R}^* \equiv p^*_T$.
Tangential components of magnetic field and velocity, on 
the other hand, may be discontinuous across any wave and
their jumps must be computed using (\ref{eq:out}) or (\ref{eq:in}).

At first sight, one might be tempted to proceed as in the 
adiabatic case, that is, by writing $\vec{U}^*$ and $\vec{F}^*$
in terms of the seven unknowns $\rho^*$, $u^*$, $v^*$, $w^*$, 
$B_y^*$, $B_z^*$ and $p_T^*$, and then using Eq. (\ref{eq:out})
and (\ref{eq:in}), together with the continuity of $\rho^*$, 
$u^*$ and $p_T^*$, to determine the jumps accordingly.
However, this approach is mathematically incorrect since
the resulting system of equations is overdetermined.
This can be better understood by explicitly writing the 
first and second component of Eq. (\ref{eq:out}):
\begin{eqnarray}
\label{eq:incorrect1}
 S_\alpha\left(\rho^* - \rho_\alpha\right) & = & 
 \rho^*u^* - \rho_\alpha u_\alpha      \,, \\ 
\label{eq:incorrect2}
 S_\alpha\left(\rho^*u^* - \rho_\alpha u_\alpha\right) & = & 
 \rho^*(u^*)^2 + p_T^* - \rho_\alpha u_\alpha -  p_{T_\alpha}  \,,
\end{eqnarray}
where I have used the fact the $\rho^*$, $u^*$ and $p^*_T$ are
constants in each of the three states.
Eqns (\ref{eq:incorrect1}) and (\ref{eq:incorrect2}) provide 
four equations (two for $\alpha=L$ and two for $\alpha=R$) 
in the three unknowns $\rho^*$, $u^*$ and $p^*_T$ and, as such,
they cannot have a solution.
Furthermore, the situation does not improve when 
the equations for the tangential components are included,
since more unknowns are brought in.

The problem can be disentangled by writing 
density and momentum jump conditions in terms
of four unknowns rather than three.
This is justified by the fact that, according
to the chosen representation, density and momentum components
of $\vec{U}^*$ and $\vec{F}^*$ are continuous across the Riemann 
fan and, in line with the consistency condition,
they should be represented by their HLL averages:
\begin{equation}\label{eq:rhostar}
  \rho^* \equiv \rho^*_L = \rho^*_c = \rho^*_R = \rho^{\rm hll}  \,,
\end{equation}
\begin{equation}\label{eq:mxstar}
  m_x^* \equiv m^*_{x_L} = m^*_{x_c} = m^*_{x_R} = m_x^{\rm hll}  \,,
\end{equation}
\begin{equation}\label{eq:velstar}
  F^*_{[\rho]} \equiv F^*_{[\rho]_L} = F^*_{[\rho]_c} = F^*_{[\rho]_R} = F_{[\rho]}^{\rm hll}  \,,
\end{equation}
\begin{equation}\label{eq:fmxstar}
 F^*_{[m_x]} \equiv F^*_{[m_x]L} = F^*_{[m_x]c} = F^*_{[m_x]R} = F^{\rm hll}_{[m_x]}  \,,
\end{equation}
where $\rho^{\rm hll}$, $m^{\rm hll}$, $F_{[\rho]}^{\rm hll}$ and 
$F^{\rm hll}_{[m_x]}$ are given by the first two components of Eq. 
(\ref{eq:uhll}) and Eq. (\ref{eq:fhll}).

This set of relations obviously satisfies the jump conditions across all the 
four waves.
The absence of the entropy mode, however, leaves the velocity in the Riemann 
fan, $u^*$, unspecified. 
This is not the case in the adiabatic case, where one has 
$m^* = F^*_{[\rho]} = \rho^*u^*$ and from the analogous consistency condition
the unique choice $\rho^{\rm hll} u^* = m_x^{\rm hll}$.
Still, in the isothermal case, one has the freedom to define 
$u^* = F^{\rm hll}_{[\rho]}/\rho^{\rm hll}$ or
$u^* = m_x^{\rm hll}/\rho^{\rm hll}$.
The two choices are not equivalent and one can show that only
the first one has the correct physical interpretation an ``advective" velocity. 
This statement becomes noticeably true in the limit of vanishing magnetic fields, 
where transverse velocities are passively advected and thus carry 
zero jump across the outermost sound waves. 
Indeed, by explicitly writing the jump conditions for the
transverse momenta, one can see that the conditions 
$v^*_\alpha=v_\alpha$ and $w^*_\alpha=w_\alpha$ can be fulfilled only 
if $\rho^* u^* = F^*_{[\rho]}$.

Thus one has, for variables and fluxes in the \emph{star} region, 
the following representation:
\begin{equation}\label{eq:star}
 \vec{U}^* = \left(\begin{array}{c}
  \rho^* \\ \noalign{\medskip}
  m^*_x    \\ \noalign{\medskip}
  \rho^*v^* \\ \noalign{\medskip} 
  \rho^*w^* \\ \noalign{\medskip} 
  B^*_y     \\ \noalign{\medskip}  
  B^*_z\end{array}\right)
 \,,\quad
 \vec{F}^* = \left(\begin{array}{c}
 \rho^*u^* \\ \noalign{\medskip}  
 F^*_{m_x} \\ \noalign{\medskip}  
 \rho^*v^*u^* - B_xB^*_y \\ \noalign{\medskip} 
 \rho^*w^*u^* - B_xB^*_z \\ \noalign{\medskip} 
  B^*_yu^*  - B_xv^*     \\ \noalign{\medskip} 
  B^*_zu^*  - B_xw^*\end{array}\right)  \,,
\end{equation}
which gives $8$ variables per state and thus a total of $24$ 
unknowns.
The 4 continuity relations (\ref{eq:rhostar})--(\ref{eq:fmxstar}) 
together with the 12 jump conditions (\ref{eq:out}) across the outer waves 
can be used to completely specify $\vec{U}^*_\alpha$ (for
$\alpha = L$ and $\alpha = R$) in terms of $\vec{U}_L$ and
$\vec{U}_R$.
Note that the total pressure does not explicitly enter in the 
definition of the momentum flux.
The remaining $4$ unknowns in the intermediate region 
($v^*_c$, $w^*_c$, $B^*_{y_c}$ and $B^*_{z_c}$) may be found 
using (\ref{eq:in}). However, one can see that 
\begin{eqnarray}
 (S^*_\alpha - u^*)\rho^* v^*_c      + B_xB^*_{y_c} & = &
 (S^*_\alpha - u^*)\rho^* v^*_\alpha + B_xB^*_{y_\alpha}   \label{eq:cv} \,, \\
 (S^*_\alpha - u^*)\rho^* w^*_c      + B_xB^*_{z_c} & = &
 (S^*_\alpha - u^*)\rho^* w^*_\alpha + B_xB^*_{z_\alpha}  \label{eq:cw} \,, \\
 (S^*_\alpha - u^*)B^*_{y_c}      + B_xv^*_{c} & = &
 (S^*_\alpha - u^*)B^*_{y_\alpha} + B_xv^*_\alpha      \label{eq:Byc} \,, \\
 (S^*_\alpha - u^*)B^*_{z_c}      + B_xw^*_{c} & = &   
 (S^*_\alpha - u^*)B^*_{z_\alpha} + B_xw^*_\alpha \label{eq:Bzc} \,,
\end{eqnarray}
cannot be simultaneously satisfied for $\alpha=L$ and $\alpha=R$
for an arbitrary definition of $S^*_\alpha$, unless the equations
are linearly dependent, in which case one finds
\begin{equation}\label{eq:sstar}
 S^*_L = u^* - \frac{|B_x|}{\sqrt{\rho^*}}\,,\quad
 S^*_R = u^* + \frac{|B_x|}{\sqrt{\rho^*}}\,.
\end{equation}
This choice uniquely specifies the speed of the rotational discontinuity in
terms of the average Alfv{\'e}n velocity $|B_x|/\sqrt{\rho^*}$.
Thus only $4$ (out of $8$) equations can be regarded independent and 
one has at disposal a total of $24$ independent equations, consistently with 
a well-posed problem.
 
Proceeding in the derivation, direct substitution of (\ref{eq:star}) into Eqns (\ref{eq:out}) 
together with (\ref{eq:rhostar}), (\ref{eq:sstar}) and the 
assumption $u^* = F^{\rm hll}_{[\rho]}/\rho^{\rm hll}$ yields 
\begin{equation}\label{eq:vstar}
 \rho^*v^*_\alpha = \rho^*v_\alpha - B_xB_{y_\alpha} 
                    \frac{u^* - u_\alpha}{(S_\alpha - S^*_L)(S_\alpha - S^*_R)} \,,
\end{equation}
\begin{equation}\label{eq:wstar}
 \rho^*w^*_\alpha = \rho^*w_\alpha - B_xB_{z_\alpha} 
                    \frac{u^* - u_\alpha}{(S_\alpha - S^*_L)(S_\alpha - S^*_R)} \,,
\end{equation}
\begin{equation}\label{eq:Bystar}
  B^*_{y_\alpha} = \frac{B_{y_\alpha}}{\rho^*}
                   \frac{\rho_\alpha(S_\alpha - u_\alpha)^2 - B_x^2}
                        {(S_\alpha - S^*_L)(S_\alpha - S^*_R)}  \,,
\end{equation}
\begin{equation}\label{eq:Bzstar}
  B^*_{z_\alpha} = \frac{B_{z_\alpha}}{\rho^*}
                   \frac{\rho_\alpha(S_\alpha - u_\alpha)^2 - B_x^2}
                        {(S_\alpha - S^*_L)(S_\alpha - S^*_R)} \,,
\end{equation}
which clearly shows that $(v^*_\alpha,w^*_\alpha)\to(v_\alpha,w_\alpha)$ 
when $\vec{B}\to\vec{0}$.
As in the adiabatic case, the consistency condition
should be used to find $v^*_c$, $w^*_c$, $B^*_{y_c}$ 
and $B^*_{z_c}$:
\begin{eqnarray}
 \rho^*v^*_c & = & \frac{(\rho^*v^*_L + \rho^*v^*_R)}{2} + \frac{X(B^*_{y_R} - B^*_{y_L})}{2}   \label{eq:vsc} \,,\\
 \rho^*w^*_c & = & \frac{(\rho^*w^*_L + \rho^*w^*_R)}{2} + \frac{X(B^*_{z_R} - B^*_{z_L})}{2}   \label{eq:wsc} \,,\\
 B^*_{y_c}   & = & \frac{(B^*_{y_L} + B^*_{y_R})}{2}   + \frac{(\rho^*v^*_R - \rho^*v^*_L)}{2X} \label{eq:Bysc}\,,\\
 B^*_{z_c}   & = & \frac{(B^*_{z_L} + B^*_{z_R})}{2}   + \frac{(\rho^*w^*_R - \rho^*w^*_L)}{2X} \label{eq:Bzsc}\,,
\end{eqnarray}
where $X = \sqrt{\rho^*}\textrm{sign}(B_x)$.
The inter-cell numerical flux $\hat{\vec{F}}$ required in Eq. (\ref{eq:oned}) 
can now be computed by sampling the solution on the $x/t=0$ axis:
\begin{equation}\label{eq:fluxfunction}
 \hat{\vec{F}} = \left\{\begin{array}{ll}
  \vec{F}_L                                & \textrm{for}\quad S_L > 0         \\ \noalign{\medskip}
  \vec{F}_L + S_L(\vec{U}^*_L - \vec{U}_L) & \textrm{for}\quad S_L < 0 < S^*_L \\ \noalign{\medskip}
  \vec{F}^*_c                              & \textrm{for}\quad S^*_L < 0 < S^*_R \\ \noalign{\medskip}
  \vec{F}_R + S_R(\vec{U}^*_R - \vec{U}_R) & \textrm{for}\quad S^*_R < 0 < S_R \\ \noalign{\medskip}
  \vec{F}_R                                & \textrm{for}\quad S_R < 0         \\ \noalign{\medskip}
\end{array}\right. \,,
\end{equation}
where $\vec{F}^*_c$ is computed from (\ref{eq:star}) using (\ref{eq:vsc}) 
through (\ref{eq:Bzsc}).

To conclude, one needs an estimate for the upper and lower signal 
velocities $S_L$ and $S_R$. I will adopt the simple estimate
given by \cite{Davis88}, for which one has
\begin{equation}
 S_L = \min\left(\lambda_{f-}(\vec{U}_L), \lambda_{f-}(\vec{U}_R)\right) \,,
 S_R = \min\left(\lambda_{f+}(\vec{U}_L), \lambda_{f+}(\vec{U}_R)\right) \,,
\end{equation}
with $\lambda_{f\pm}(\vec{U})$ given by the first of (\ref{eq:eigenvalues}).
Other estimates have been proposed, see for example \cite{EMRS91}.
  
The degenerate cases are handled as follows:
\begin{itemize}
 \item for zero tangential components of the magnetic field and 
       $B^2_x > a^2\rho_\alpha$, a degeneracy occurs where 
       $S^*_\alpha \to S_\alpha$ and the denominator in 
       Eqns. (\ref{eq:vstar})--(\ref{eq:Bzstar}) vanishes. 
       In this case, one proceeds as in the adiabatic case by imposing
       transverse components of velocity and magnetic field to have 
       zero jump;
 \item the case $B_x\to 0$ does not poses any serious difficulty and only
       the states $\vec{U}^*_L$ and $\vec{U}^*_R$ have to be evaluated
       in order to compute the fluxes.
       One should realize, however, that the final representation of the solution 
       differs from the adiabatic counterpart. 
       In the full solution, in fact, Alfv{\'e}n and slow modes all degenerate into a 
       single intermediate tangential discontinuity with speed $u^*$.
       In the adiabatic case, this degeneracy can still be well described
       by the presence of the middle wave, across which only normal 
       velocity and total pressure are continuous.
       Although this property is recovered by the adiabatic HLLD solver, 
       the same does not occur in our formulation because of the initial
       assumption of constant density through the Riemann fan.
       The unresolved density jump, in fact, can be thought of as the 
       limiting case of two slow shocks merging into a single wave.
       Since slow shocks are not considered in the solution, density will 
       still be given by the single HLL-averaged state.
       Tangential components of velocity and magnetic field, on the other 
       hand, can yet be discontinuous across the middle wave.
       Thus, in this limit, the solver does not reduce to the single
       state HLL solver.
\end{itemize}

This completes the description of the proposed isothermal HLLD Riemann solver. 
The suggested formulation satisfies the consistency conditions 
by construction and defines a well-posed problem, 
since the number of unknowns is perfectly balanced 
by the number of available equations:  
$12$ equations for the outermost waves,
$8$ continuity conditions (for $\rho^*$, $u^*$, $F^*_{[\rho]}$, 
$F^*_{[m_x]}$ across the inner waves) and the $4$ equations 
for the tangential components. 
The positivity of the solver is trivial since no energy equation is present
and the density is given by the HLL-averaged state (\ref{eq:rhostar}). 

\section{Numerical Tests}
\label{sec:num}
%
%
%

The performance of the new Riemann solver is now investigated through 
a series of selected numerical tests in one and two dimensions.
One dimensional shock tubes (in \S\ref{sec:sod}) consist in a 
decaying discontinuity initially separating two constant states. 
The outcoming wave pattern comprises several waves and
an analytical or reference solution can usually be obtained 
with a high degree of accuracy. 
For this reason, one dimensional tests are commonly used to check the 
ability of the code in reproducing the correct structure.

The decay of standing Alfv{\'e}n waves in two dimensions is consider 
in \S\ref{sec:alfv}. The purpose of this test is to quantify the amount of 
intrinsic numerical dissipation inherent to a numerical scheme. 
A two dimensional tests involving the propagation of a 
blast wave in a strongly magnetized environment is investigated
in \S\ref{sec:blast} and finally, in \S\ref{sec:orstang}, 
the isothermal Orszag-Tang vortex system and its transition 
to supersonic turbulence is discussed.

Eq. (\ref{eq:1st_ord}) is used for the spatially and temporally
first order scheme. 
Extension to second order is achieved with the $2^{\rm nd}$ order
fully unsplit Runge-Kutta Total Variation Diminishing (TVD) of 
\cite{GS96} and the harmonic slope limiter of \cite{vLeer74}.
For the multidimensional tests, the magnetic field is evolved using 
the constrained transport method of \cite{B99}, where staggered
magnetic fields are updated, e.g. during the predictor step, 
according to
\begin{eqnarray}
  B^{n+1}_{x,i+\HALF,j} &  =  &B^{n}_{x,i+\HALF,j} - \frac{\Delta t}{\Delta y_j}
  \left(\Omega^z_{i+\HALF,j+\HALF} - \Omega^z_{i+\HALF,j-\HALF}\right) \,,
  \\ \noalign{\medskip} 
  B^{n+1}_{y,i,j+\HALF} &  =  &B^{n}_{y,i,j+\HALF} + \frac{\Delta t}{\Delta x_i}
  \left(\Omega^z_{i+\HALF,j+\HALF} - \Omega^z_{i-\HALF,j+\HALF}\right) \,.
\end{eqnarray}
The electric field $\Omega^z_{i+\HALF,j+\HALF}$ is computed by 
a spatial average the upwind fluxes available at zone interfaces 
during the Riemann solver step:
\begin{equation}
  \Omega^z_{i+\HALF,j+\HALF} = 
 \frac{- \hat{F}^{[B_y]}_{i+\HALF,j} - \hat{F}^{[B_y]}_{i+\HALF,j+1} 
       + \hat{F}^{[B_x]}_{i,j+\HALF} + \hat{F}^{[B_x]}_{i+1,j+\HALF}}{4}\,,
\end{equation}
where $\hat{F}^{[B_y]}$ is the $B_y$ component of the flux
(\ref{eq:fluxfunction}) available during the sweep along the 
x-direction. A similar procedure applies to $\hat{F}^{[B_x]}$.
 

\subsection{One Dimensional Shock Tubes}
\label{sec:sod}
%

\begin{table}\begin{center}
\begin{tabular*}{0.5\textwidth}%
     {@{\extracolsep{\fill}}ccccccccc}
 Test &  $\rho$ & $u$  & $v$  & $w$ & $H_x$ & $H_y$ & $H_z$ &  $t_s$\\ \hline\hline
1L    &   1     & 0    &  0   &  0  &  3    &  5    &  0    &   0.1 \\
1R    &   0.1   & 0    &  0   &  0  &  3    &  2    &  0    &    -  \\ \hline
2L    &   1.08  & 1.2  & 0.01 & 0.5 &  2    &  3.6  &  2    &   0.2\\
2R    &   1     & 0    &  0   &  0  &  2    &  4    &  2    &    - \\ \hline
3L    &   0.1   &  5   &  0   &  0  &  0    &  -1  &  -2    &   0.25\\
3R    &   0.1   & -5   &  0   &  0  &  0    &   1  &   2    &     -  \\ \hline
\end{tabular*}
\caption{Initial conditions for the one-dimensional shock tube problems
         presented in the text. Here $(H_x, H_y, H_z) = \sqrt{4\pi}(B_x, B_y, B_z)$.
         The final integration time ($t_s$) is given in the last column.}
\label{tab:ic}
\end{center}\end{table}

An interface separating two constant states
is placed in the middle of the domain $[0,1]$ at $x=0.5$.
Unless otherwise stated, the isothermal speed of sound $a$ is 
set to unity. 
States to the left and to the right of the 
discontinuity together with the final integration time
are given in Table \ref{tab:ic}. 
The reader may also refer to \cite{KRJH99} for the first 
and second problems and to \cite{B98b} for the third one.
For the sake of comparison, the linearized Riemann solver
of Roe (properly adapted to the isothermal case from \cite{CG97})
and the simple single-state HLL solver of \cite{HLL83} 
will also be used in the computations.
In order to better highlight the differences between the selected 
Riemann solvers, integrations are carried with a spatially and 
temporally first order-accurate scheme on $400$ uniform zones.
The Courant number is set to $0.8$ in all calculations.

\begin{figure}\begin{center}
 \includegraphics[width=0.75\textwidth]{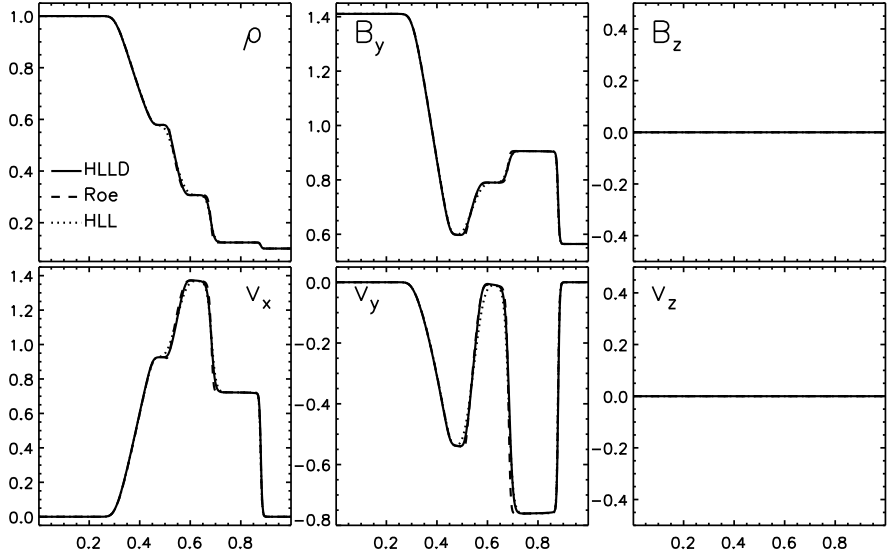}
 \caption{Computed profiles for the first shock tube problem 
          at time $t=0.1$. The top panels show, from 
          left to right, density ($\rho$), and transverse components
          of magnetic field ($B_y$ and $B_z$, respectively). 
          Shown on the bottom panels are the three
          components of velocity.
          Solid, dashed and dotted lines refer to computations 
          carried with the HLLD, Roe and HLL Riemann solvers,
          respectively.}
 \label{fig:sod1}
\end{center}\end{figure}%
Profiles of density, velocities and magnetic fields for the first 
shock tube are shown in Fig. (\ref{fig:sod1}) .
Solid, dashed and dotted lines refer to computations carried out
with the HLLD, Roe and HLL solvers, respectively. 
The decay of the initial discontinuity results in the 
formation, from left to right, of a fast and a slow rarefaction 
waves followed by a slow and a fast shocks. No rotational discontinuity 
is present in the solution.
The leftmost rarefaction fan and the rightmost fast shock 
are equally resolved by all three solvers.
The resolution of the slow rarefaction is essentially 
the same for the Roe and the HLLD schemes, but is smeared out
over more computational zones for the HLL method.
At the slow shock, differences become less evident although 
the solution obtained with the HLLD solver seems to spread 
the discontinuity over a slightly higher number of zones with 
respect to the Roe scheme.
Although the HLLD solver should, in principle, offer no
gain over the simpler HLL method in absence of rotational 
waves, the wave patterns is still reproduced more accurately 
than the HLL scheme. 

\begin{figure}\begin{center}
 \includegraphics[width=0.75\textwidth]{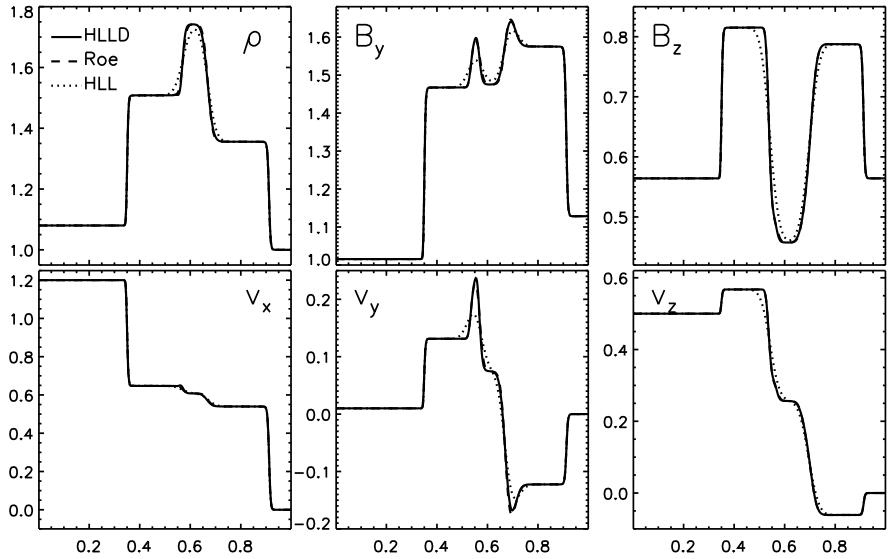}
 \caption{Computed profiles for the second shock tube problem 
          at time $t=0.2$. The top panels show, from 
          left to right, density ($\rho$), $y$ and $z$ components
          of magnetic field ($B_y$ and $B_z$, respectively). 
          Shown on the bottom panels are the three
          components of velocity.
          Solid, dashed and dotted lines refer to computations 
          carried with the HLLD, Roe and HLL Riemann solvers,
          respectively.}
 \label{fig:sod2}
\end{center}\end{figure}
\begin{figure}\begin{center}
 \includegraphics[width=0.75\textwidth]{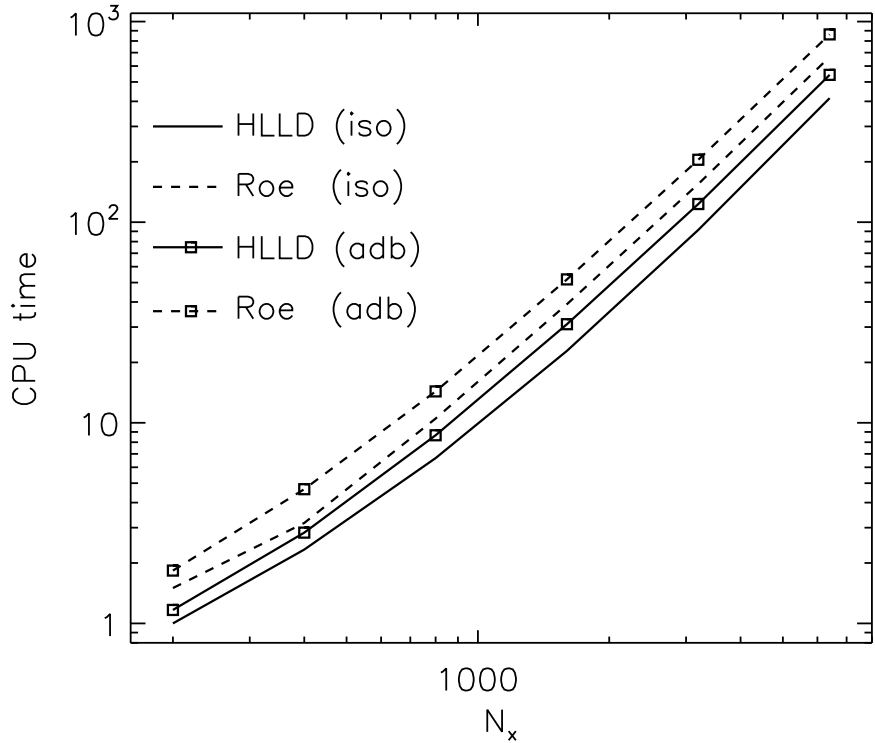}
 \caption{CPU time (in units of the fastest run) for the 
          HLLD (solid lines) and Roe schemes (dashed lines).
          The adiabatic cases (with $\Gamma=1.01$) 
          are overplotted using boxes.}
 \label{fig:adb_vs_iso}
\end{center}\end{figure}

The ability to resolve Alfv{\'e}n waves is better observed in 
the second shock tube (test 2) with initial conditions given in Table \ref{tab:ic}.
In this case one has the formation of three pairs of waves
which, from the exterior to the interior of the domain, can be identified 
with a pair of fast magnetosonic shocks, a pair of rotational discontinuities
and a pair of slow shock waves, see Fig. \ref{fig:sod2}.
One can note the poorer performance of the HLL solver in resolving
the rotational discontinuities which are smeared out on several 
zones. Likewise, the peak in the $y$ component of magnetic field 
between the Alfv{\`e}n wave and the slow shock is considerably 
underestimated ($\gtrsim 4\%$).
On the contrary, the Roe and HLLD schemes give comparable 
performances in terms of reduced numerical diffusion on essentially 
all waves, including the two slow shocks.
The plots shown in Fig \ref{fig:sod2} show, in fact, almost
complete overlap.
In terms of efficiency, however, the new solver offers ease
of implementation over the Roe solver which requires 
the characteristic decomposition of the Jacobian matrix.
This is better quantified in Fig. \ref{fig:adb_vs_iso}, 
where the two schemes are compared in terms of CPU execution times
as functions of the mesh size.
At the resolutions employed ($200$ to $6400$), the isothermal 
HLLD is faster than the Roe solver by a factor $\sim 35 \div  40\%$.
The same performance is found if one compares the efficiencies of 
the two solvers by running the same test with an adiabatic equation 
of state (by setting $\Gamma=1.01$).
Furthermore, direct inspection of Fig. \ref{fig:adb_vs_iso} 
reveals that at lower resolution (i.e. $\lesssim 600$) 
the CPU time ratio adiabatic/isothermal is $\sim 1.2$, whereas
it increases up to $\sim 1.35$ at higher resolutions.
This justifies the need for codes specifically designed for
IMHD, as already anticipated in \S\ref{sec:intro}.


\begin{figure}\begin{center}
 \includegraphics[width=0.75\textwidth]{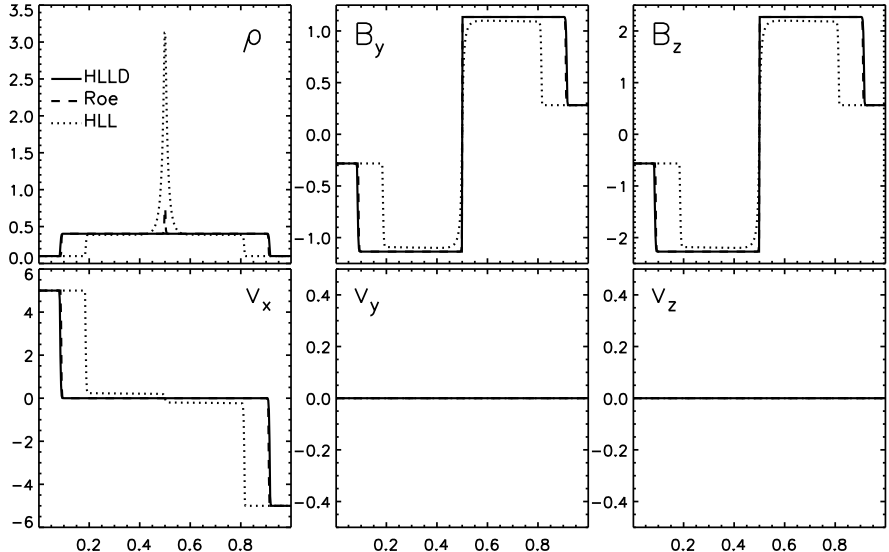}
 \caption{Computed profiles for the third shock tube problem 
          at time $t=0.25$. The top panels show, from 
          left to right, density ($\rho$), $y$ and $z$ components
          of magnetic field ($B_y$ and $B_z$, respectively). 
          Shown on the bottom panels are the three
          components of velocity.
          Solid, dashed and dotted lines refer, respectively, to 
          computations carried with the HLLD, Roe and HLL Riemann solvers.}
 \label{fig:sod5}
\end{center}\end{figure}
The collision of oppositely moving high Mach number flows is
examined in the third shock tube, see also \cite{B98b}. 
This problem also illustrates the performances of the new
solver in the limit of vanishing normal component of 
the magnetic field, $B_x = 0$, as discussed at the end
of \S\ref{sec:solver}.
The wave pattern is bounded by two outermost fast shocks
traveling in opposite directions and a stationary tangential 
discontinuity (in the middle) carrying no jump in total pressure 
and normal velocity, see Fig \ref{fig:sod5}.    
The two shocks propagate at the same speed which can be found 
analytically by solving the jump conditions (\ref{eq:out}) 
with respect to $S_\alpha$, the upstream density and magnetic field 
(the upstream velocity is zero).
This yields the value $|S_\alpha| \approx 1.6529$ and therefore
predicted shock positions at $\approx 0.5 \pm 0.4132$ when $t=0.25$.
At the resolution of $400$ zones, the HLL solver performs 
noticeably worse than the other two schemes, resulting
in slower shock propagation speeds, hence the underestimated 
jumps.
On the contrary, the HLLD and Roe solvers perform very similarly 
across all three waves and are able to reproduce the expected 
shock position with relative errors less than $10^{-2}$.
The spurious density spike at the center of the grid is a 
numerical artifact generated by the solver, which is trying to compensate the 
reduced magnetic pressure (caused by the spreading of $B_y$ and $B_z$ over
a region of finite width across the discontinuity) by increasing the density 
in such a way that the total pressure is maintained constant.
This effect is manifestly more pronounced for the HLL scheme yielding 
a peak value of $\approx 3.1$ and become less noticeable for the Roe
scheme ($\approx 0.73$). The HLLD solver does not exhibit this feature
and yields a flat density profile, concordantly with the exact 
solution.

\begin{figure}\begin{center}
 \includegraphics[width=0.8\textwidth]{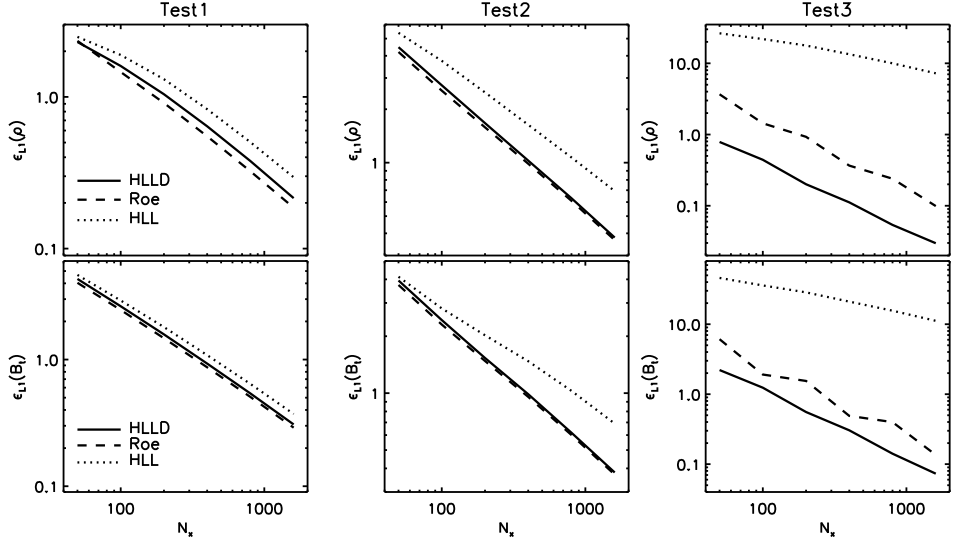}
 \caption{Percent L1 norm errors (computed as 
          $\epsilon_{L_1}(q) = 10^2/N_x\sum_{i=1}^{i=N_x} |q^{\rm ex}(x_i) - q(x_i)|$
          where $q^{\rm ex}$ and $q$ are the exact and numerical solutions) as 
          function of the resolution, 
          for the first (left panels), second (middle panels) and 
          third (right panels) shock tube problems.  
          Top and bottom rows give the error in density and transverse 
          magnetic field $B_y$, respectively.}
 \label{fig:errors}
\end{center}\end{figure}

In order to quantify the errors and check the convergence 
properties of the scheme, computations have been repeated at 
different grid sizes $N_x =50,100,200,400,800,1600$ and the 
$L_1$-norm errors have been measured against a reference solution.
The reference solution was obtained with the second order
version of the scheme using an adaptive grid with $3200$ 
coarse zones and $4$ levels of refinement, therefore corresponding to 
an effective resolution of $512,000$ uniform zones.
Results shown in Fig. \ref{fig:errors} confirm that the
new solver reproduces the correct solution with considerably
smaller errors than the simple HLL scheme.
In the second test problem, the Roe and HLLD schemes 
perform nearly the same with density and transverse magnetic field
errors being almost halved ($\lesssim 4\%$, at the highest resolution)
with respect to the HLL method.
In the third problem, the HLLD solver is able to capture the solution 
with even higher accuracy than the Roe method indeed yielding, 
at the highest resolution, errors of $\approx 0.03\%$ and 
$\approx 0.07\%$ in density and magnetic field (respectively)
compared with $0.1\%$ and $0.13\%$ obtained with the Roe scheme. 
The unusually higher errors ($\gtrsim 10\%$) observed in the
solutions obtained with the HLL scheme are accounted for 
by the higher density overshoot occurring at the center of the grid.

\subsection{Decay of Standing Alfv{\'e}n Waves}
\label{sec:alfv}
%
%

The amount of numerical viscosity and/or resistivity intrinsic
to computations carried on discrete grids may be estimated with 
a simple numerical test involving the decay of standing Alfv{\'e}n waves,
\cite{RJF95,KRJH99}.
In a fluid with non-zero dynamic shear viscosity $\mu$ and 
resistivity $\eta$, waves are characterized by a decay rate given 
by 
\begin{equation}
  \Gamma_A = \frac{1}{2}\left(\frac{\mu}{\rho_0} + \eta\right) |\vec{k}|^2\,,
\end{equation}
where $\rho_0$ is the background density and $\vec{k}$ is the wave number
vector.
The inevitable approximations introduced by any numerical scheme lead 
to an effective intrinsic numerical viscosity (and resistivity) that
can be also accounted for by the particular Riemann solver being used.
Thus, a measure of $\Gamma_A$ can be directly related to the 
``effective" $\mu$ and $\eta$ inherent to the scheme.

\begin{figure}\begin{center}
 \includegraphics[width=0.45\textwidth]{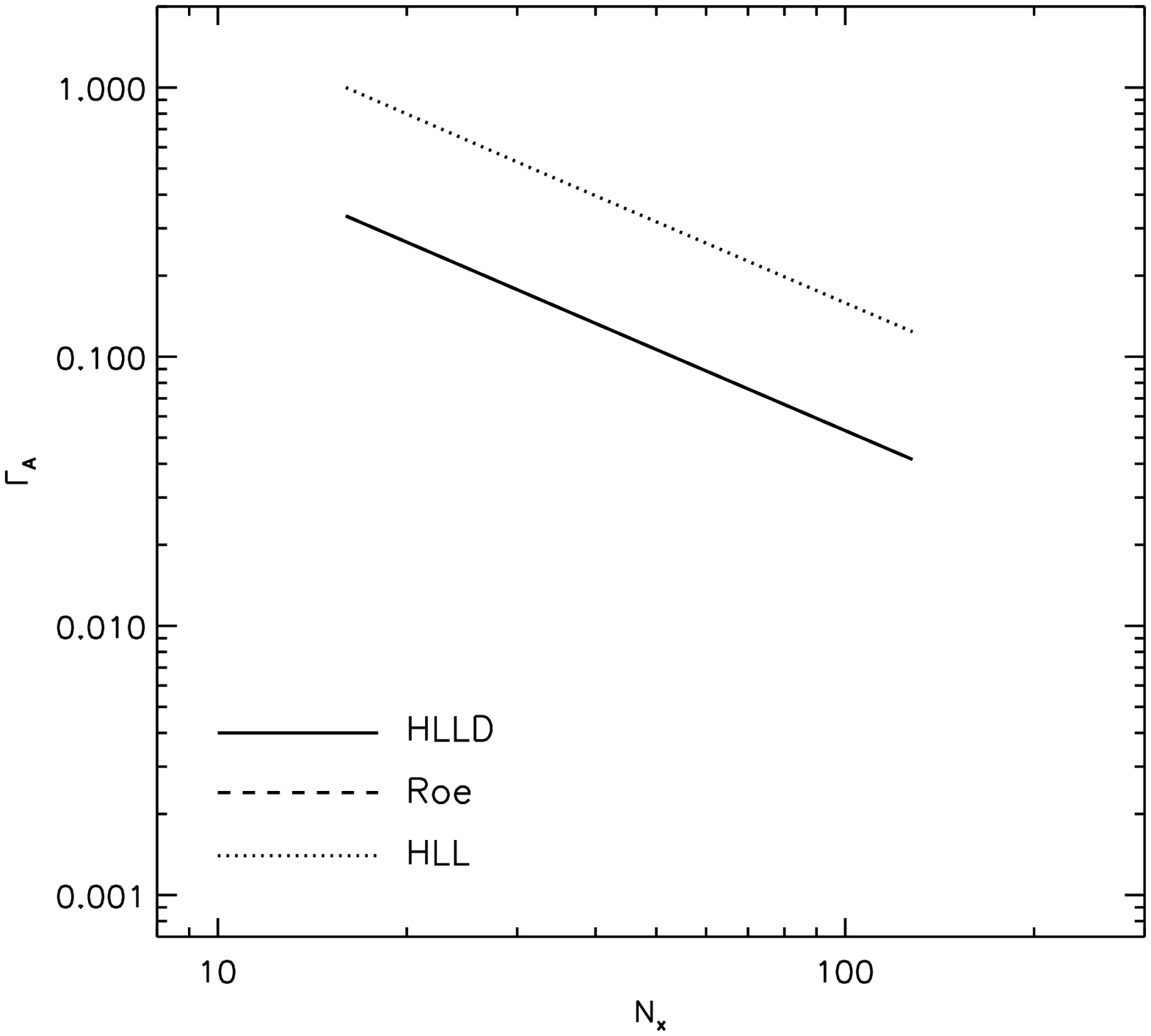}%
 \includegraphics[width=0.45\textwidth]{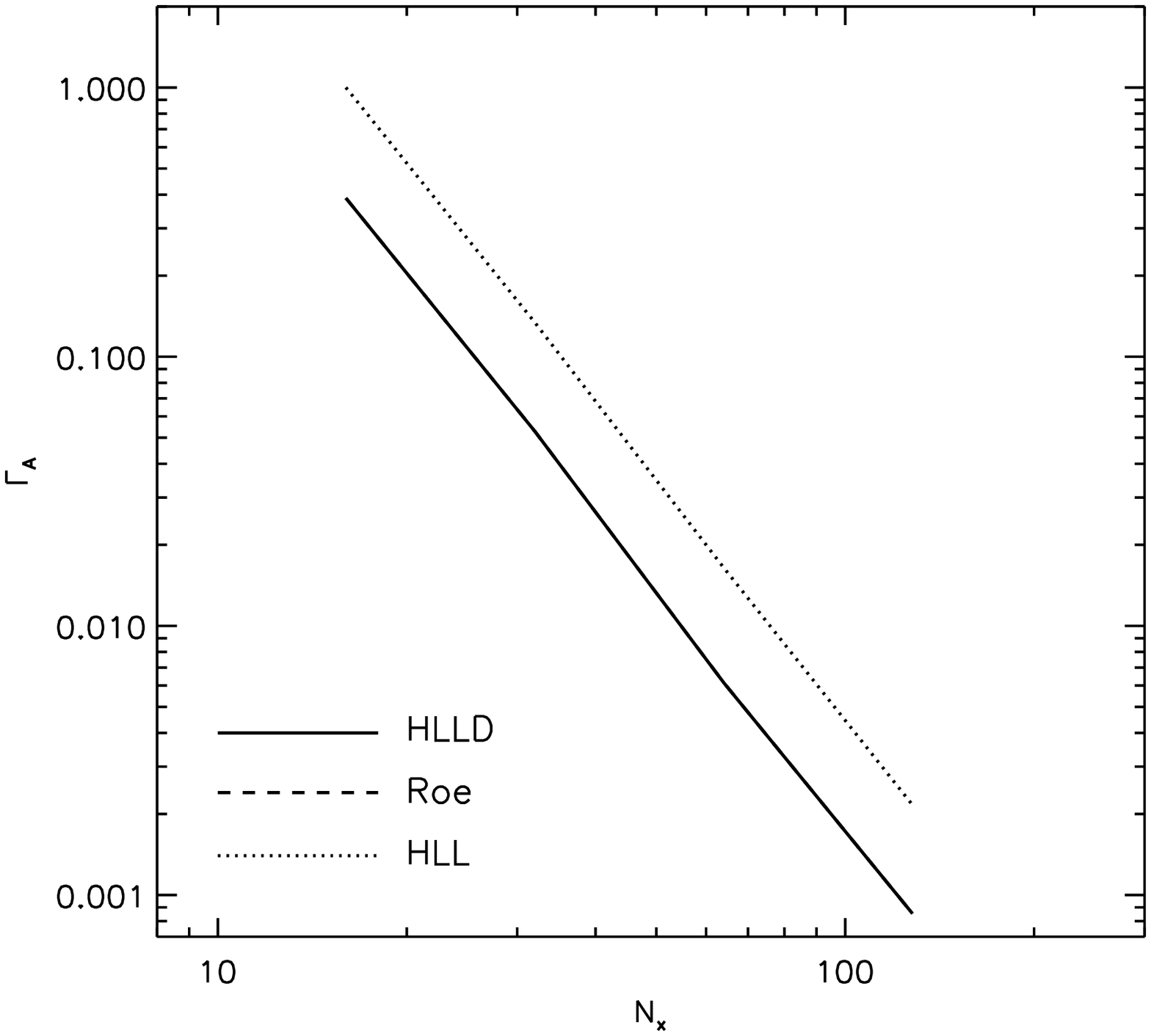}
 \caption{Decay rate as function of resolution for the 
          standing Alfv{\'e}n wave test. Left and right panels
          give the computed decay rate on a log-log scale for 
          the first and the second order scheme, respectively.
          At a given resolution, the decay rate is measured 
          by first fitting the peaks $\delta B_z^{\max}(t)$ of 
          $\delta B_z(t) = \sqrt{\int\int B^2_z(t) dxdy}$
          with respect to time, and then by approximating  
          the corresponding $\Gamma_A$ with 
          $d\log\left(\delta B_z^{\max}(t)\right)/dt$ 
          computed as average slope over the time integration interval.
          Note that the HLLD and Roe solvers yields essentially the 
          same decay rate and the two lines cannot be distinguished.}
 \label{fig:alfv}
\end{center}\end{figure}

The initial condition consists of a uniform background medium with 
constant density $\rho = 1$ and magnetic field 
$\vec{B} = \hat{\vec{x}}$.
The $x$ and $y$ components of velocity are set to zero everywhere
and the sound speed is $a=1$.
Standing Alfv{\'e}n waves are set up along the main diagonal 
by prescribing
\begin{equation}
 v_z = \epsilon c_A\sin\left(\vec{k}\cdot\vec{x}\right)
\end{equation}
where $c_A = 1/\sqrt{2}$, $\vec{k} = 2\pi(\hat{\vec{x}} + \hat{\vec{y}})$ 
and $\epsilon = 0.1$.
Computations are carried over the unit square $[0,1]^2$ until 
$t=10$ at different resolutions ($16^2$, $32^2$, $64^2$, $128^2$ zones)
and for the three Riemann solvers.
Periodic boundary conditions are imposed on all sides.
Following \cite{KRJH99}, the decay rate corresponding to a given
resolution (for $0\le t\le 10$) has been derived by fitting the 
peaks of the r.m.s. of the $z$ component of the magnetic field, 
$\delta B_z(t)$, with respect to time.
Fig. \ref{fig:alfv} shows the normalized decay rates for the first 
and second order schemes at different resolutions and for
the selected Riemann solvers.
The amount of numerical viscosity intrinsic to the HLLD and Roe scheme
turns out to be virtually the same, whereas it is a factor of $3$ (for
the $1^{\rm st}$ order) and $\approx 2.6$ (for the $2^{\rm nd}$ order)
higher for the HLL solver.
The slope of the curves is the same for all of the methods, therefore 
confirming the selected integration order of accuracy ($\propto N^{-1}_x$ or 
$\propto N^{-2}_x$ for the $1^{\rm st}$ or $2^{\rm nd}$ order scheme,
respectively).

\subsection{Cylindrical Blast Wave}
\label{sec:blast}
%
%

\begin{figure}\begin{center}
 \includegraphics[width=0.75\textwidth]{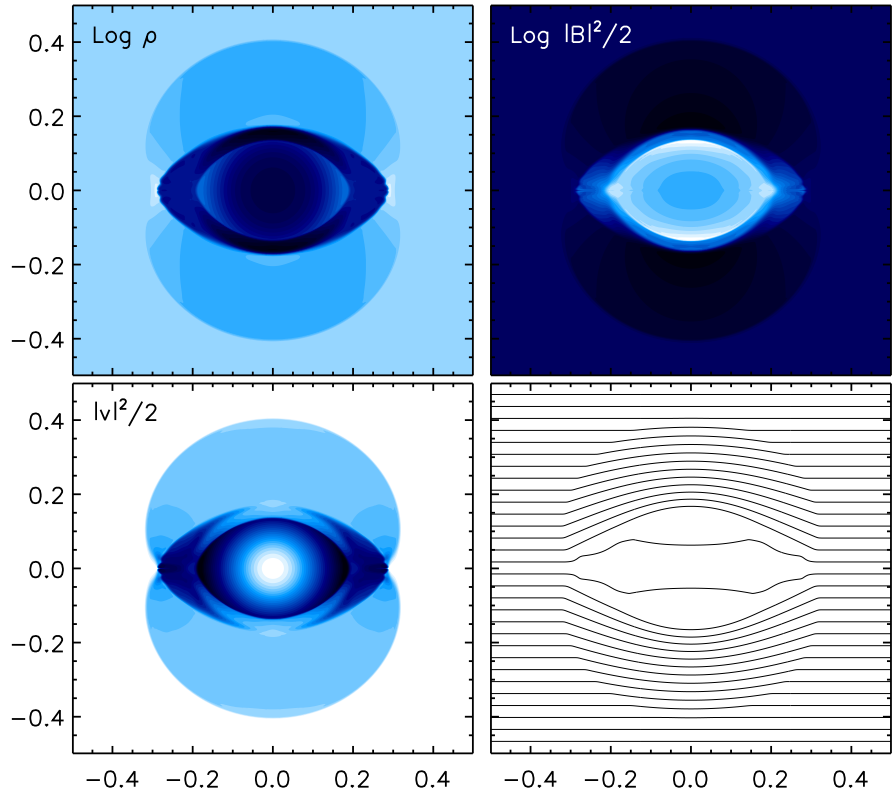}
  \caption{Colored contours for the blast wave problem at $t=0.09$.
           Density logarithm (upper left panel), magnetic pressure
           logarithm (upper right), specific kinetic energy (bottom
           left) and magnetic field lines (bottom right) are shown.}
 \label{fig:blast}
\end{center}\end{figure}
A uniform medium with constant density $\rho=1$ and
magnetic field $\vec{B} = 10/\sqrt{4\pi}\hat{\vec{i}}$ fills 
the square domain $[-\HALF,\HALF]^2$.
An over-pressurized region ($p = 100$) is delimited by a small
circle of radius $0.05$ centered around the origin.
The speed of sound is set to unity everywhere and free-flow boundary
conditions are imposed on every side of the domain.

Cylindrical explosions in Cartesian coordinates provide a useful 
benchmark particularly convenient in checking the robustness of the 
code and the response of the algorithm to different kinds of degeneracies. 
Fig. \ref{fig:blast} shows the results obtained 
on $401^2$ uniform zones at $t=0.09$.
The ratio of thermal and magnetic pressure
is $\approx 0.251$ thus making the explosion anisotropic.
The outer region is delimited by an almost cylindrical fast forward shock 
reaching its maximum strength on the $y$ axis (where the 
magnetic field is tangential) and progressively weakening as $y\to 0$.
In the central region, density and magnetic fields are gradually 
depleted by a cylindrical rarefaction wave enclosed by an eye-shaped
reverse fast shock bounding the inner edge of a higher density 
ring.
The outer edge of this ring marks another discontinuity which 
degenerates into a tangential one on the $y$ axis and into 
a pure gas dynamical shock at the equator ($y=0$) where $B_y=0$.
The HLLD solver provides adequate and sharp resolution of all 
the discontinuities and the result favorably compares to the 
integrations carried out with the other Riemann solvers (not shown
here) and with the results of \cite{B98b}.

The computational execution times for the HLL, HLLD and Roe solvers 
stay in the ratio $1:1.21:2.01$, thus showing that the new method 
is approximately $20\%$ more expensive whereas the Roe solver 
results in computations which are almost twice as slow.

\subsection{Isothermal Orszag Tang Vortex}
\label{sec:orstang}
%
%
  
%
\begin{figure}\begin{center}
 \includegraphics[width=0.75\textwidth]{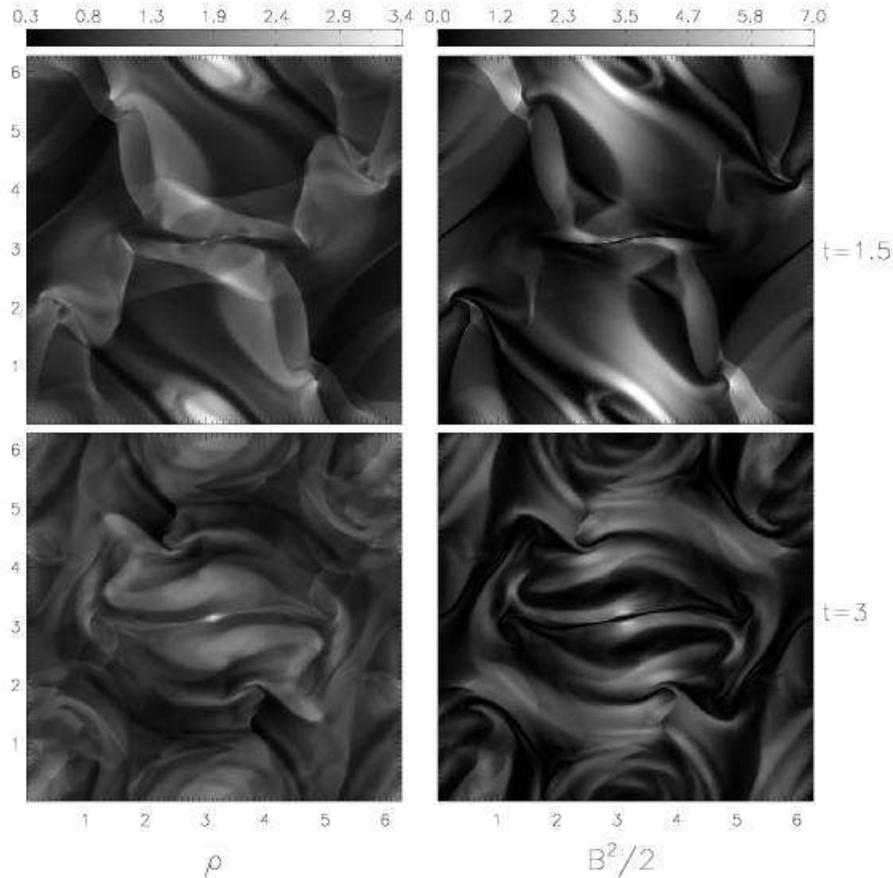}
  \caption{Evolution of the Orszag-Tang vortex at $t=1.5$ (top) and
           $t=3$ (bottom). Density and magnetic pressure are shown 
           on the left and on the right, respectively.
           Note the formation of a magnetic island in the center at
           $t = 3$.}
 \label{fig:orstang}
\end{center}\end{figure}
The compressible Orszag-Tang vortex system describes 
a doubly periodic fluid configuration undergoing 
supersonic MHD turbulence in two dimensions.
Although an analytical solution is not known, its simple and
reproducible set of initial conditions has made it a 
widespread benchmark for inter-scheme comparison, see for
example \cite{Toth00}.
An isothermal version of the problem was previously considered by 
\cite{B98b}. In his initial conditions, however, the numerical value 
of the sound speed was not specified. 
For this reason, I will consider a somewhat different
initial condition with velocity and magnetic fields prescribed as
\begin{equation}
 \vec{v} = a\left(-\sin y\hat{\vec{x}} + \sin x\hat{\vec{y}}\right) \,,
\end{equation}  
\begin{equation}
 \vec{B} = B_0\left(-\sin y\hat{\vec{x}} + \sin (2x)\hat{\vec{y}}\right) \,.
\end{equation}  
The value of the speed of sound is $a=2$ and the magnetic field amplitude
is $B_0 = a\sqrt{3/5}$. 
The density is initialized to one all over. This configuration 
yields a ratio of thermal to magnetic pressure of $10/3$, as
in the adiabatic case. The domain is the box $[0,2\pi]^2$ with 
periodic boundary conditions imposed on all sides.
Fig. \ref{fig:orstang} shows the density and magnetic pressure
distributions at $t=1.5$ and $t=3$. 
The initial vorticity distribution spins the fluid clockwise
and density perturbations steepen into shocks around 
$t\approx 0.7$. Because of the imposed double periodicity, 
shocks re-emerge on the opposite sides of the domain and 
broad regions of compression are formed.
The dynamics is regulated by multiple interactions of 
shock waves, which lead to the formation of small scale 
vortices and density fluctuations.
The shocks continuously affect the spatial distributions 
of density and magnetic field.
Magnetic energy is dissipated through current sheets
convoyed by shocks at the center of the domain where 
reconnection takes place around $t\approx 1.5$, through the ``Y" point.
By $t=3$ the kinetic energy has been reduced by more
than $50 \%$ and shocks have been weakened by the
repeated expansions, leaving small scale density 
fluctuations. 

Note that a magnetic island featuring a high density spot forms at the center
of the domain by $t=3$. This corresponds to a change in the magnetic field 
topology where an ``O" point forms in the central region.
It should be emphasized that the same feature is observed when the Roe 
solver is employed during the integration and it is also distinctly 
recognizable in the results of \cite{B98b}.
However, it requires at least twice the resolution to be seen if the 
fluxes are computed with the HLL solver. 
Yet, the magnetic island does not form at sufficiently lower resolution
for any solver.
Although its origin has to do with the complex phenomenology of MHD 
reconnection which lies outside the scope of this paper, it is speculated 
that its presence may be numerically induced by a decreased effective 
resistivity across the central current sheet. Future studies should
specifically investigate on this issue.

The total execution times for the three solvers (HLL, HLLD and Roe) 
scale as $1:1.25:2.06$, thus confirming the results obtained
for the previous test problem.

\section{Conclusions}
\label{sec:conclusions}
%
%
%

A new Riemann solver for the equations of magnetohydrodynamics
with an isothermal equation of state has been derived.
The solver leans on a multi-state Harten-Lav-van Leer representation of the
solution inside the Riemann fan, where only fast and rotational 
discontinuities are considered. 
Under this approximation, density, normal momentum and the corresponding 
flux components inside the solution are given by their HLL averages.
This is equivalent to the assumption that normal velocity, density and 
total pressure may be regarded constant across the solution.
It has been shown that this consistently leads to a three-state representation
where only tangential vectors may be discontinuous across the inner 
rotational waves. 
The solver is simple to implement and does not require a characteristic
decomposition of the Jacobian matrix.

The performances and efficiency of the new method have been demonstrated through 
one dimensional shock tube problems as well as multidimensional 
problems.
In terms of accuracy, the proposed method of solution greatly 
improves over the traditional single-state HLL approach in that it provides
adequate and superior resolution of rotational discontinuities and 
a reduced numerical diffusion.
Furthermore, it has been found that the new ``HLLD" solver yields accuracies 
comparable to or even better than the Roe solver by retaining simplicity and 
greater ease of implementation over the latter.
In terms of efficiency, the proposed HLLD solver is only $\approx 20-25\%$
slower than the HLL method but considerably faster (up to $\approx 60\%$
in some cases) than the Roe scheme.
    

The author whishes to thank the support and hospitality received at
the center of magnetic self-organization at the University of Chicago.



\end{document}